# Reverberation measurement of the inner radius of the dust torus in NGC 4151 during 2008-2013

V. L. Oknyansky[1], N. V. Metlova[1], O. G. Taranova[1], V. I. Shenavrin[1], B. P. Artamonov[1],

C. M. Gaskell[2], Di-Fu Guo[3]

[1] Sternberg Astronomical Institute, Moscow M.V. Lomonosov State University, Moscow, Russian Federation, *oknyan@mail.ru*

[2] Department of Astronomy and Astrophysics, University of California, Santa Cruz, USA

[3] School of Space Science and Physics, Shandong University, Weihai, China

ABSTRACT. We investigate the correlation between infrared (***JHKL***) and optical (***B***) fluxes of the variable nucleus of the Seyfert galaxy NGC 4151 using partially published data for the last 6 years (2008-2013.). Here we are using the same data as in Oknyansky et al. (2014), but include also optical (***B***) data from Guo et al. We find that the lag of flux in all the infrared bands is the same, 40 ± 6 days, to within the measurement accuracy. Variability in the ***J*** and ***K*** bands is not quite simultaneous, perhaps due to the differing contributions of the accretion disk in these bands. The lag found for the ***K*** band compared with the ***B*** band is not significantly different from earlier values obtained for the period 2000-2007. However, finding approximately the same lags in all IR bands for 2008-2013 differs from previous results at earlier epochs when the lag increased with increasing wavelength. Examples of almost the same lag in different IR bands are known for some other active nuclei. In the case of NGC 4151 it appears that the relative lags between the IR bands may be different in different years. The available data, unfortunately, do not allow us to investigate a possible change in the lags during the test interval. We discuss our results in the framework of the standard model where the variable infrared radiation is mainly due to thermal re-emission from the part of the dusty torus closest to the central source. There is also a contribution of some IR emission from the accretion disk, and this contribution increases with decreasing wavelength.

Some cosmological applications of obtained results are discussed.

**Key words:** AGN, optical and IR variability, time delay, data analysis, dust torus, cosmology.

## 1. Introduction

NGC4151 is one of the most popular and well studied AGNs. it is most bright and high variable object, which is very often used as an typical Sy1. The generally accepted model of an AGN is a supermassive black hole surrounded by an accretion disk (AD). The AD emits mostly soft X-ray, UV, optical continuum, and also gives some contribution to the near infrared. Above the AD dense clouds emit the broad emission lines. Outside this broad-line region is the narrow-line region. Variable near-IR radiation is associated with the region at a radius between the radii of the BLR and the NLR. This region is usually associated with the part of the optically-thick dusty torus closest to the central source (Hönig and Kishimoto 2011). The presence of such a torus is the key to explaining of the observed differences in the spectra of type 1 and type 2 Seyfert nuclei by the torus eclipsing our direct view of the continuum radiation and the broad lines. It is also believed that the dusty torus radiates in the infrared, as a result of heating by shorter wavelength radiation from the central regions of the AD. Closer to the centre the dust is completely (or largely) sublimated and delayed infrared variability gives us the estimate of radius of the "dust holes" around the central source (Oknyanskij and Horne 2001), i.e. the radius of the region where the dust is absent. The first measurement of the inner radius of the dust torus in NGC 4151 based on the cross-correlation analysis of IR and optical variability was obtained Oknyanskij (1993). Then, it was reviled that the inner radius varies depending on the luminosity of the central source (Oknyansky et al. 1989). However, these changes do not occur simultaneously, so as to restore the dust after the high luminosity states takes time. For NGC 4151 this time is at least a few years (Oknyanskij, 2006, 2008).

This work is a continuation of our series of papers in which we measure the inner radius of the dust torus in NGC 4151 and other AGNs from the delay of the variability in the near infrared to the optical (see details and references at Oknyansky et al 2014) . Despite the significant growth of theoretical and observational studies of AGN in the IR, our knowledge of the dust, its origin and the detailed morphology of these objects remain largely incomplete.

### 2. Method of cross-correlation analysis

The inner radius of the dust torus in AGNs can be measured on the basis of cross-correlation analysis of near-infrared and optical variability. Cross-correlation analysis of astronomical time series is complicated because the sampling is usually uneven. As a rule, series of astronomical observations inevitably have gaps because of seasons when the object is invisible and interruptions due to the full moon, weather, and observing schedules. Classic cross-correlation analysis methods were developed only for uniformly sampled time series. The analysis of non-uniform astronomical series requires special techniques. At the present work we are using the method MCCF (see details and references at Oknyansky et al., 2014). At the heart of our method is the ICCF method (Gaskell and Spark, 1986), but we strive to reduce the contribution of the interpolation errors introducing certain limit *MAX* interval used for the interpolated points. We use only those interpolated points which are separated in time from the nearest observation points by no more than the value *MAX*.

### 3. Observational data

In this paper, we use the same observational data of IR and optical photometry for the interval 2008-2013, as in Oknyansky (2014), but supplemented it with optical *B* data by Guo et al. (2014). These additional CCD observations were obtained on 15 nights using the 1.0-m telescope at Weihai Observatory of Shandong University (observed from 2009May to 2013 February). The data were reduced to the same system as in Oknyansky et al. (2014). The formal accuracy of the photoelectric and CCD measurements is not worse than 1-2%, but there may be systematic differences between measurements obtained on different instruments. We estimated that these errors do not exceed 10%. Our light curves in the *JHKL* infrared band and in the optical *B*-band are shown in Fig.1.

### 4. Cross-correlation analysis.

Cross-correlation functions MCCF for the infrared *JHKL* bands versus the *B* band for 2008-2013 are presented in Fig. 2. It can be seen that the main peak for all of these cross-correlation functions in 36-44 days. The size of the lag between the *K* band and the optical remained almost the same as the lag we found for the period 2000-2007 (Oknyanskij et. al. 2008 ), but the time delay for variability at *L* band before was significantly bigger - 105±5 days.

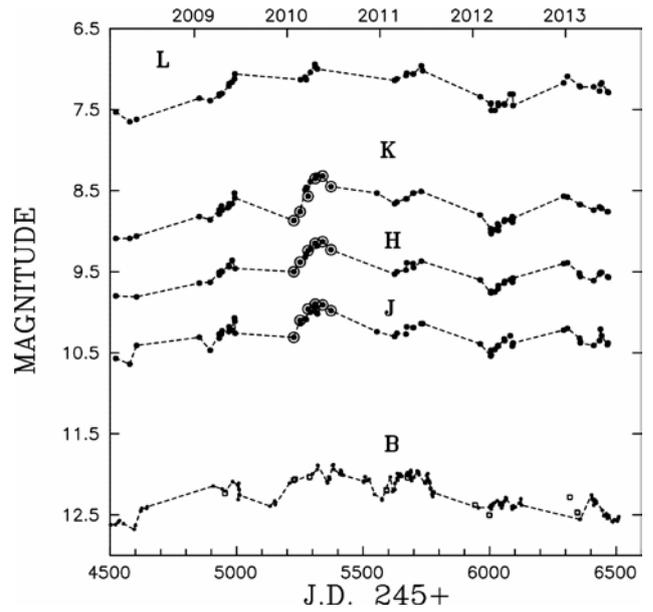

Figure 1. Infrared (*JHKL* bands) and optical (*B*-band) light curves for 2008-2013. For the IR light curves the filled circles are our data and the open circles are the data of Schnülle et al. (2013) reduced to the same system. For the *B*-band light curves the points are as follows: small filled circles – photoelectric and CCD measurements (our data and Roberts and Rumstey (2012), see details at Oknyansky et al. (2014)); open boxes – CCD photometry Guo et al. (2014) reduced to the same system .

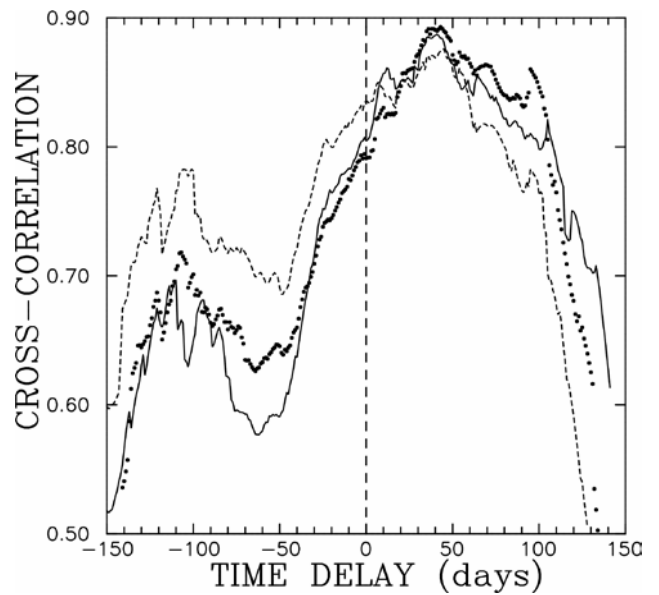

Figure 2. Cross-correlation functions calculated using the MCCF method with *MAX* = 30 days for *J* (dashed line), *K* (dots), *L* (solid line), versus *B* for 2008-2013.

## 5. Cosmological applications

The ability to measure cosmological constants on the basis of the of delay of near-infrared variability relative to the optic was first mentioned by Kobayashi et al. (1998) and was independently proposed Oknyanskij (1999, 2002). At the same time it was first shown (Oknyanskij 1999, 2002; Oknyanskij and Horne 2001) that for a number of AGN the time delay of variability (in the $K$-band) depends on the luminosity in the UV as $r_s \sim (L_{UV})^{1/2}$ in accordance with the theoretical result Barvainis (1987, 1992). In the recent publications Yoshii et al. (2014) and Hönig (2014) considered in details the possibility of measuring the cosmological constants on the basis of a study of delays in the IR variability. Of particular interest is the ability to measure distances to objects with large redshifts. For the case it is necessary to take into account the possible dependence of the delays from the wavelength in the IR range to make any appropriate reduction. The equation for such a reduction was derived by Oknyansky and Horne (2001) on the basis of the theory (Barvanis 1987). However, IR delay may be not depending on the wavelength for the some part of the AGN, as well as it can be a temporary feature of a particular object, as it turned out in the case of NGC 4151. Accordingly, there is not exist any common rule for all objects to correct delays for redshift, in dispite to suggested by Oknyanskij and Horne (2001) and by Yoshii et al. (2014). Therefore, it is desirable to study variability of AGNs in several IR bands.

The presence of variability in AGNs is necessary for this method of measuring cosmological distances, although it is a particular problem, as the inner radius of the dust torus does not change in sync with changes in luminosity. Accordingly, the long-term photometric studies of objects is requested. Possible to measure the inner radius of a dust torus by infrared interferometry, but unfortunately it can be difficult or impossible for AGNs with high redshifts.

## 6. Conclusion

We firstly found that lags in different IR bands in NGC 4151 are approximately the same. This has been observed previously in some other AGNs, but for NGC 4151 it seems that it may be a temporary feature.

Using IR lags for cosmological applications is a very promising method, but it needs detailed investigations of optical and IR variability of selected AGNs as well as improved theoretical models.

*Acknowledgements*. This work was supported by the Russian Foundation for Basic Research (grant nos. 14-02-01274,13-02-00136, and 12-02-01237)

**References**

Barvainis, R.: 1987, *Ap. J.* **320**,537.
Barvainis, R.:1992 *Ap. J.* **400**, 502.
Guo D.F. et al.: 2014, *Res. in A&A.*, **14**, 923.
Gaskell, C. M; Sparke, L. S.:1986, *Ap. J.* **305**, 175.
Hönig S.F. ; Kishimoto M. 2011: *A&A*,. **524**, A121.
Hönig S.F.: 2014, *Ap. J.Lett.*, **774**, L4.
Kobayashi Y. et. al.:1998, *Proc. SPIE*, **3352**, 120.
Oknyansky V.L. et al.: 2014, *Astron. Lett.*,**40**, 527.
Oknyanskij V.L. et al.: 2008, OAP, **21**, 79.
Oknyanskij, V. L.; Horne, K.:2001, *ASP Conf. Proc.* **224**, 149.
Oknyanskij V.L. et.al.:2006, *ASP Conf. Ser.* **360**, 75.
Oknyanskij, V.L:1993 *Astron. Lett.* **19**, 416 .
Oknyanskij V.L. et.al.:1999, *Astron. Lett.* **25**, 483.
Oknyanskij V.L.:2002, *ASP Conf. Proc.*, **282**, 330.
Roberts C.A., Rumstey, K. R.:2012 *J. South. Associat. Res. Astron.* **6**, 47.
Schnülle A. et al.: 2013, *A&A, 557,* L13.
Yoshii Y. et al.:2014, *Ap.J.Lett.*, **784**, L11.